\documentclass[aps,prb,preprint,groupedaddress]{revtex4-2}
\usepackage{natbib}
\usepackage{setting_prb}

\newif\ifcalc
\calcfalse

\begin{document}
%\preprint{}

\title{Transport through a monolayer-tube junction:\\ 
sheet-to-tube spin current in silicene}

\author{Yuma Kitagawa,$^{1,2}$ Yuta Suzuki,$^{1,2}$ Shin-ichiro Tezuka,$^{2}$ and Hiroshi Akera$^{3}$}

\affiliation{
$^{1}$Division of Applied Physics, Graduate School of Engineering, Hokkaido University, Sapporo, Hokkaido, 060-8628, Japan\\
$^{2}$Sensing Research \& Development Department, Innovation Center, Marketing Headquarters, Yokogawa Electric Corporation, Tokyo, 180-8750, Japan\\
$^{3}$Division of Applied Physics, Faculty of Engineering, Hokkaido University, Sapporo, Hokkaido, 060-8628, Japan
}

\date{\today}

\begin{abstract}
A method is developed to calculate the electron flow between an atomic monolayer sheet and a tube 
with use of tunneling matrix elements between monolayer sheets 
and applied 
to the spin current from monolayer silicene with sublattice-staggered current-induced spin polarization 
to silicene tube. 
Calculated sheet-to-tube spin current exhibits an oscillation as a function of the tube circumferential length since 
the Fermi points in the tube cross the Fermi circle in the sheet. 
It is also shown that the spin current with spin in the out-of-plane direction, 
which is absent in the sheet-sheet junction (including twisted sheets) with the $C_3$ rotational symmetry, 
appears in an oscillating form owing to the broken $C_3$ symmetry in the tube-sheet junction.
\end{abstract}
%\tableofcontents

%\keywords{}

\maketitle

%%%%%%%%%%%%%%%%%%%%%%%%%%%%%%%%%%%%%%%%%%%%%%%%%%%%%%%%%%%%%
\section{Introduction}\label{sec_Introduction}
%%%%%%%%%%%%%%%%%%%%%%%%%%%%%%%%%%%%%%%%%%%%%%%%%%%%%%%%%%%%%
Symmetry of the structure strongly affects the transport. 
As a textbook example, the current is parallel to the electric field in a crystal with cubic symmetry, 
while the current direction can deviate from the field direction in a general crystal structure \cite{Ashcroft_Mermin1976solid}. 
In atomic layers with the van der Waals interaction, 
flexible layer stacking can manipulate the symmetry to control transport properties \cite{Gorbachev2014, Offidani2017, veneri_twist_2022, lee_charge_spin_2022, Kurebayashi2022}. 
As a typical example, 
the twist of bilayer graphene 
%\textcolor{red}{[twisted bilayer grapheneの提案、理論的記述、作製、chirality-induced特性の理論・実験に関する文献]}
\cite{lopesdossantos_graphene_2007, bistritzer_transport_2010, suarez_morell_flat_2010, bistritzer_moire_2011, cao_superlattice_2016, cao_unconventional_2018, yananose_chirality_induced_2021} 
breaks the mirror symmetry to make the bilayer chiral. 
We have studied in a recent paper \cite{Kitagawa2023} the interlayer spin current in twisted bilayer silicene 
generated by the sublattice-staggered current-induced spin polarization (CISP) 
\cite{yanase_magneto-electric_2014, Zelezny2014, wadley_electrical_2016, Watanabe2018, suzuki_spin_2023} 
in the lower layer of bilayer silicene  
and shown that 
the twist, by breaking the mirror symmetry, gives rise to  
the component of the interlayer spin current with spin in the direction rotated in-plane by 90 degrees 
from the CISP direction 
in addition to that in the CISP direction. 
On the other hand, the $C_3$ symmetry is preserved in a twisted bilayer 
which consists of monolayers with the $C_3$ symmetry. 
Since the spin current with the out-of-plane spin direction is absent 
due to the $C_3$ symmetry as derived in our previous paper \cite{Kitagawa2023}, 
we expect that it appears by breaking the $C_3$ symmetry. 

In this paper we theoretically study changes in the spin transport 
when the $C_3$ symmetry is broken by replacing the upper layer of twisted bilayer silicene with a tube. 
In this junction of the lower sheet and the tube 
we calculate the spin current from the monolayer sheet with the CISP to the tube 
as a function of the tube circumferential length. 
To calculate the spin current through this junction, 
we derive an approximate formula for 
tunneling matrix elements of a tube-sheet junction for an arbitrary lattice structure in each of the tube and the sheet. 
This formula is expressed by 
corresponding matrix elements of the sheet-sheet junction, 
for which the formula has been derived 
in previous theories \cite{bistritzer_transport_2010, bistritzer_moire_2011, koshino_interlayer_2015} 
for an arbitrary bilayer and expressed with interlayer hopping integrals between atoms. 
With use of the formula for matrix elements of the tube-sheet junction, 
we derive the formula for the electron flow through the junction, 
which is used to calculate the spin current through the tube-sheet junction of silicene. 
The formula we derive for tunneling matrix elements and the electron flow 
can be used to study tube-sheet junctions formed from 
an arbitrary combination of atomic monolayers 
such as graphene, hexagonal boron nitride, phosphorene, and transition-metal dichalcogenides.  

This paper is organized as follows.
In Sec.\,\ref{sec_calculation_method} 
we derive a formula for tunneling matrix elements and that for the electron flow 
through a tube-sheet junction formed by arbitrary atomic monolayers. 
In Sec.\,\ref{sec_silicene_junction} 
we calculate the spin current from a silicene monolayer sheet to a silicene tube 
by using the formula derived in Sec.\,\ref{sec_calculation_method} 
and by solving the Boltzmann equation for the electron distribution in the sheet with the CISP  
in the relaxation-time approximation. 
Conclusions are given in Sec.\,\ref{sec_conclusion}.

%==============================================================================================
%\iffigure
\begin{figure}[ht]\centering
	\includegraphics[width=90mm]{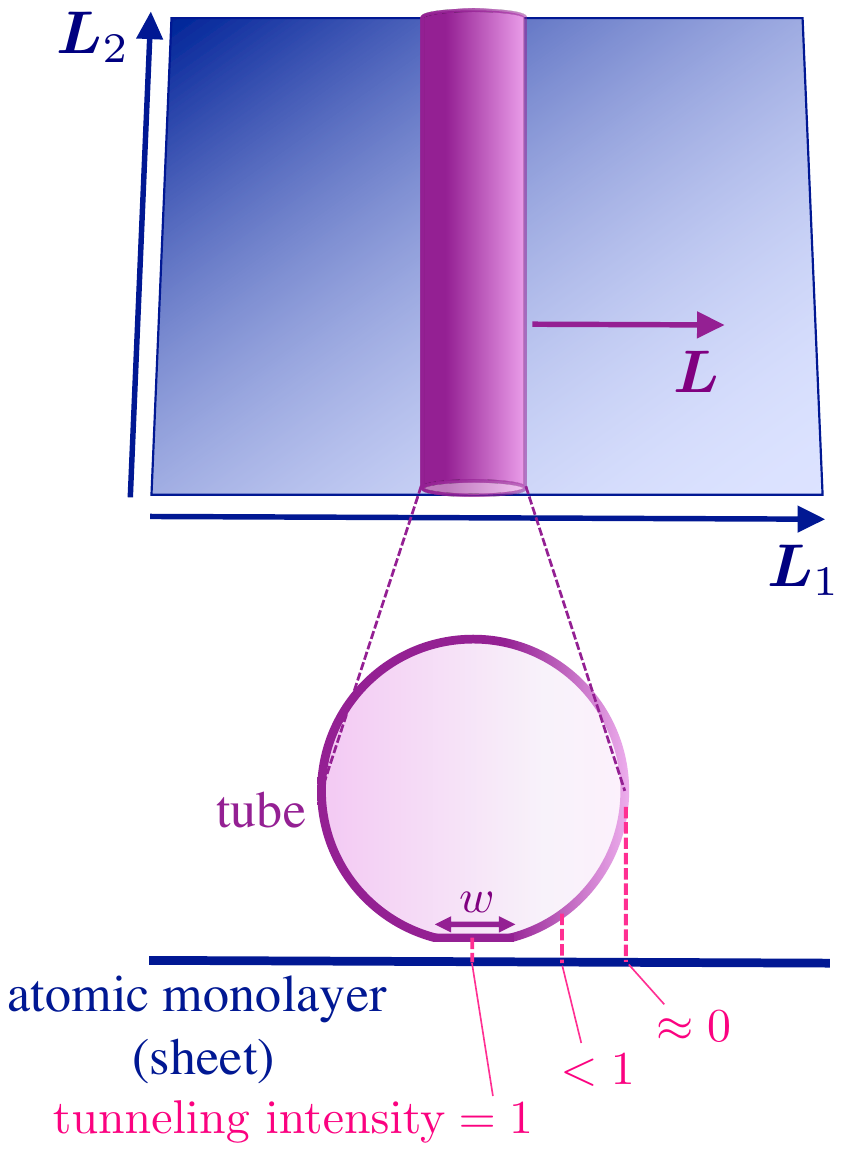}
	\vskip -4.5cm
	\caption{A junction of an atomic monolayer and a tube. 
	$\vL$ is the chiral vector of the tube. 
	We take $\vL_1$ in the direction of $\vL$ and $\vL_2$ parallel to the tube axis.
	$w$ is the width of the junction.}
	\label{fig_tube_sheet_junction}
\end{figure}
%\fi
%==============================================================================================

%%%%%%%%%%%%%%%%%%%%%%%%%%%%%%%%%%%%%%%%%%%%%%%%%%%%%%%%%%%%%
\section{Calculation method for tube-sheet junction}\label{sec_calculation_method}
%%%%%%%%%%%%%%%%%%%%%%%%%%%%%%%%%%%%%%%%%%%%%%%%%%%%%%%%%%%%%
\subsection{Tunneling matrix elements between an atomic monolayer and a tube}
%%%%%%%%%%%%%%%%%%%%%%%%%%%%%%%%%%%%%%%%%%%%%%%%%%%%%%%%%%%%%%
In this section we develop an approximate method which can  
express tunneling matrix elements between an atomic monolayer sheet and a tube 
by those between monolayer sheets. 
This approximation is applicable to tube-sheet junctions with junction width larger than the Fermi wavelength of the tube. 

We start with expressing tunneling matrix elements between monolayer sheets. 
We assume that each monolayer has the translational  symmetry described by 
primitive translation vectors, 
$\va_1^l$ and $\va_2^l$ in the lower sheet and 
$\va_1^u$ and $\va_2^u$ in the upper sheet. 
Two monolayers may have different crystal structures. 
We express eigenvectors of each monolayer in a linear combination of atomic basis vectors 
$\ket{\vR_\lambda^\alpha X \sigma}$ 
where $\vR_\lambda^\alpha$ is the position vector of the $\lambda$th atom in the unit cell of layer $\alpha\, (=l,u)$ and 
different vectors in each atom are labelled by 
orbital index $X\, (= 1s,\, 2s,\, 2p_x,\, 2p_y,\, 2p_z,\, \cdots)$ and spin $\sigma\, (= \uparrow,\, \downarrow)$. 
Crystal basis vectors are expressed by the sum of atomic basis vectors over $N_s$ unit cells of each sheet 
%-------------------------------------------------------------------------------------------------------------------------
\begin{alignat}{99}\begin{split}\label{eq_bloch_basis_vector}
	\ket{\alpha\vk \lambda X \sigma} &= \frac{1}{\sqrt{N_s}}\sum_{\vR_\lambda^\alpha} e^{i\vk \cdot\vR_\lambda^\alpha} \ket{\vR_\lambda^\alpha X \sigma},
\end{split}\end{alignat}
%-------------------------------------------------------------------------------------------------------------------------
where $\vk$ is the two-dimensional Bloch wave vector.  
We apply the periodic boundary condition so that $\vk \cdot \vL_1$ and $\vk \cdot \vL_2$ are integers multiplied by $2\pi$ 
where $\vL_1$ and $\vL_2$ define the area of the sheet [Fig.\,\ref{fig_tube_sheet_junction}].

Then each eigenvector $\ket{\alpha n\vk}$ of the unperturbed Hamiltonian $H_0$, which satisfies
%-------------------------------------------------------------------------------------------------------------------------
\begin{alignat}{99}\begin{split}
	H_0 \ket{\alpha n\vk} &= \veps_{n\vk}^{\alpha} \ket{\alpha n\vk} ,
\end{split}\end{alignat}
%-------------------------------------------------------------------------------------------------------------------------
with $\veps_{n\vk}^{\alpha}$ the corresponding eigenenergy,
is expressed with expansion coefficients $C_{\vk\lambda X\sigma}^{\alpha n}$ by
%-------------------------------------------------------------------------------------------------------------------------
\begin{alignat}{99}\begin{split}\label{eq_sheet_eigenvector}
	\ket{\alpha n\vk} &= \sum_{\lambda X\sigma} C_{\vk\lambda X\sigma}^{\alpha n} \ket{\alpha\vk\lambda X\sigma} .
\end{split}\end{alignat}
%-------------------------------------------------------------------------------------------------------------------------
Here $n$ denotes the band index which includes the spin degree of freedom.

We consider the perturbation $\HT$ which describes the electron tunneling between monolayers. 
With $\HT$ included, the total Hamiltonian is 
%-------------------------------------------------------------------------------------------------------------------------
\begin{alignat}{99}\begin{split}
	H = H_0 + \HT .
\end{split}\end{alignat}
%-------------------------------------------------------------------------------------------------------------------------
The matrix element describing the tunneling between 
$\ket{ln\vk}$ an eigenstate in the lower sheet and 
$\ket{un'\vk'}$ that in the upper sheet is given by
$\mel{ln\vk}{\HT}{un'\vk'}$, 
the formula of which has been derived 
in previous theories \cite{bistritzer_transport_2010, bistritzer_moire_2011, koshino_interlayer_2015}  
for an arbitrary bilayer and expressed with use of interlayer hopping integrals between atoms. 

Now we make a tube from the upper sheet [Fig.\,\ref{fig_tube_sheet_junction}] so that 
an atom at $\vR_\lambda^u$ in the upper sheet overlaps that at $\vR_\lambda^u + \vL$ 
where $\vL$ is called the chiral vector and $\vL= m_1 \va_1^u + m_2 \va_2^u$ with $m_1$ and $m_2$ integers. 
We consider a tube having a large radius \cite{Ando2005} in which we can neglect the curvature effect. 
Then the tube eigenvector can be expressed with atomic basis vectors in the upper sheet plane 
and is given by 
%-------------------------------------------------------------------------------------------------------------------------
\begin{alignat}{99}\begin{split}
	\ket{tn\vkappa}_u &= \sqrt{\frac{N_s}{N_t}} \ket{un\vkappa} ,
\end{split}\end{alignat}
%-------------------------------------------------------------------------------------------------------------------------
using the sheet eigenvector $\ket{un\vkappa}$ [Eq.\,\eqref{eq_sheet_eigenvector}]. 
Unlike the basis vector $\ket{\alpha\vk \lambda X \sigma}$ [Eq.\,\eqref{eq_bloch_basis_vector}] 
used to express the sheet eigenvector $\ket{un\vk}$, 
the sum in the basis vector $\ket{\alpha\vkappa \lambda X \sigma}$ of $\ket{tn\vkappa}_u$ 
is restricted to tube atoms whose number is denoted by $N_t$.
Here the momentum $\vkappa$ is quantized in the direction of $\vL$ to be  
$\vkappa \cdot \vL/(2\pi)=\ $integers.  
The subscript $u$ of $\ket{tn\vkappa}_u$ indicates that it is expressed using basis vectors of the upper sheet.  
We denote the corresponding tube eigenvector in the three-dimensional space 
by $\ket{tn\vkappa}$ without the subscript. 

Using $\ket{tn\vkappa}_u$ we approximately express the matrix element of the tunneling between 
a lower-sheet eigenstate $\ket{ln\vk}$ and a tube eigenstate $\ket{tn'\vkappa}$ 
by 
%-------------------------------------------------------------------------------------------------------------------------
\begin{alignat}{99}\begin{split}
	\mel{ln\vk}{\HT}{tn'\vkappa} &= \mel{ln\vk}{\HT \PT}{tn'\vkappa}_u ,
\end{split}\end{alignat}
%-------------------------------------------------------------------------------------------------------------------------
with  
%-------------------------------------------------------------------------------------------------------------------------
\begin{alignat}{99}\begin{split}\label{eq_projection_to_junction}
	\PT = \sum_{\vR}t(\vR)\dyad{\vR},
\end{split}\end{alignat}
%-------------------------------------------------------------------------------------------------------------------------
where the sum is taken over position vectors $\vR$ of all atoms in the upper sheet.
The projection operator $\PT$ multiplies each atomic basis vector in $\ket{tn'\vkappa}_u$ by an intensity $t(\vR)$. 
This intensity expresses the tunneling intensity at each tube atom. 
As shown in Fig.\,\ref{fig_tube_sheet_junction}, 
tube atoms on the upper sheet plane have the full tunneling intensity of $t(\vR)=1$, 
while atoms away from the plane have a weaker intensity of $t(\vR)<1$. 
This approximation only takes into account the interatomic distance between the lower sheet and the tube 
and neglects the tube curvature which modifies the angle between atomic orbitals involved in the tunneling. 

We express the tunneling intensity $t(\vR)$ in the Fourier expansion
%-------------------------------------------------------------------------------------------------------------------------
\begin{alignat}{99}\begin{split}\label{eq_g_Fourier}
	t(\vR) = \sum_\vq \hat t(\vq) e^{i\vq\cdot\vR} ,
\end{split}\end{alignat}
%-------------------------------------------------------------------------------------------------------------------------
where $\vq \cdot \vL_1$ and $\vq \cdot \vL_2$ are integers multiplied by $2\pi$. 
Then the tunneling matrix element becomes 
%-------------------------------------------------------------------------------------------------------------------------
\begin{alignat}{99}\begin{split}
	\mel{ln\vk}{\HT}{tn'\vkappa} &= \sqrt{\frac{N_s}{N_t}} \sum_\vq \hat t(\vq) \mel{ln\vk}{\HT}{un'\vkappa +\vq (\vkappa)} ,
\end{split}\end{alignat}
%-------------------------------------------------------------------------------------------------------------------------
where
%-------------------------------------------------------------------------------------------------------------------------
\begin{alignat}{99}\begin{split}
\ket{un'\vkappa +\vq (\vkappa)} = \sum_{\lambda X\sigma} C_{\vkappa\lambda X\sigma}^{un'} \ket{u\vkappa +\vq\lambda X\sigma},
\end{split}\end{alignat}
%-------------------------------------------------------------------------------------------------------------------------
which is a vector obtained by replacing $\ket{u\vkappa\lambda X\sigma}$ in the eigenvector $\ket{un'\vkappa}$ 
by $\ket{u\vkappa +\vq\lambda X\sigma}$. 
Here we assume that 
the width $w$ of the tube-sheet junction with $t(\vR) \approx 1$ is much larger than the Fermi wavelength $\lambda_\rF^t$ of the tube. 
Since the distribution width of the Fourier coefficient $\hat t(\vq)$ is $\sim\! 2\pi/w$ 
and $|\vkappa|\approx 2\pi/\lambda_\rF^t$ for $\vkappa$ relevant to the transport between the tube and the sheet 
in the low-temperature region such that $\kB T \ll \veps_\rF^t$ ($\kB$: the Boltzmann constant, $\veps_\rF^t$: the Fermi energy of the tube), 
we have $|\vq| \ll |\vkappa|$ and $C_{\vkappa\lambda X\sigma}^{un'} \approx C_{\vkappa + \vq \lambda X\sigma}^{un'}$. 
Since $C_{\vkappa\lambda X\sigma}^{un'} \approx C_{\vkappa + \vq \lambda X\sigma}^{un'}$ leads to $\ket{un'\vkappa +\vq (\vkappa)} \approx \ket{un'\vkappa +\vq}$,  
we finally obtain
%-------------------------------------------------------------------------------------------------------------------------
\begin{alignat}{99}\begin{split}\label{eq_mel_tube_sheet}
	\mel{ln\vk}{\HT}{tn'\vkappa} &= \sqrt{\frac{N_s}{N_t}} \sum_\vq \hat t(\vq) \mel{ln\vk}{\HT}{un'\vkappa +\vq} ,
\end{split}\end{alignat}
%-------------------------------------------------------------------------------------------------------------------------
which relates the tunneling matrix element of a tube-sheet junction 
to that of a sheet-sheet junction. 

%%%%%%%%%%%%%%%%%%%%%%%%%%%%%%%%%%%%%%%%%%%%%%%%%%%%%%%%%%%%%
\subsection{Expression for the electron flow using tunneling matrix elements}
%%%%%%%%%%%%%%%%%%%%%%%%%%%%%%%%%%%%%%%%%%%%%%%%%%%%%%%%%%%%%
\subsubsection{Electron flow between atomic monolayers}
%%%%%%%%%%%%%%%%%%%%%%%%%%%%%%%%%%%%%%%%%%%%%%%%%%%%%%%%%%%%%%
As a preparation for deriving the electron flow from an atomic monolayer sheet to a tube, 
we derive the electron flow from the lower sheet to the upper one. 
Both sheets occupy a two-dimensional square space 
whose boundaries are parallel to $\vL_1$ and $\vL_2$ and have the length of $|\vL_1| = |\vL_2| =L_s$. 
We start with the number of electrons in the lower sheet 
projected onto the spin direction $\pm\gamma$ ($\gamma=x,y,z$), defined by
%-------------------------------------------------------------------------------------------------------------------------
\begin{alignat}{99}\begin{split}
	N_{\pm\gamma}^l 
	&= \tr(\rho P_{\pm\gamma} P_l), 
\end{split}\end{alignat}
%-------------------------------------------------------------------------------------------------------------------------
with $\rho$ the density operator. 
The projection operator onto the lower monolayer is defined by 
%-------------------------------------------------------------------------------------------------------------------------
\begin{alignat}{99}\begin{split}
	P_l 
	&= \sum_{n}\sum_{\vk \in B_l}\dyad{ln\vk}, 
\end{split}\end{alignat}
%-------------------------------------------------------------------------------------------------------------------------
where 
the sum with respect to $\vk$ is taken over $\vk$ satisfying the periodic boundary conditions, 
$\vk \cdot \vL_1/(2\pi)=\ $integers and $\vk \cdot \vL_2/(2\pi)=\ $integers, 
within the Brillouin zone of the lower monolayer $B_l$.  
The projection operator onto the $\pm\gamma$ spin direction is defined by  
%-------------------------------------------------------------------------------------------------------------------------
\begin{alignat}{99}\begin{split}
	P_{\pm\gamma}
	&= \dyad{\pm\gamma}, 
\end{split}\end{alignat}
%-------------------------------------------------------------------------------------------------------------------------
where 
%-------------------------------------------------------------------------------------------------------------------------
\begin{alignat}{99}\begin{split}
\sigma_{\gamma} \ket{\pm\gamma} = \pm \ket{\pm\gamma},
\end{split}\end{alignat}
%-------------------------------------------------------------------------------------------------------------------------
with $\sigma_{\gamma}$ the Pauli spin operator. 
Electrons in the lower sheet with the $\pm\gamma$ spin direction flow out to the upper sheet with the rate of 
%-------------------------------------------------------------------------------------------------------------------------
\begin{alignat}{99}\begin{split}
	J_{\pm\gamma}^{l\rightarrow u} 
	&= - \dv{N_{\pm\gamma}^l}{t} = - \tr(\dv{\rho}{t}P_{\pm\gamma} P_l) \\
	&= - \sum_{n}\sum_{\vk \in B_l} \mel**{l n\vk}{\dv{\rho}{t}P_{\pm\gamma}}{ln\vk}.
\end{split}\end{alignat}
%-------------------------------------------------------------------------------------------------------------------------
The spin current with spin in the $\gamma$ direction is given by
%-------------------------------------------------------------------------------------------------------------------------
\begin{alignat}{99}\begin{split}
	J_{\rs \gamma}^{l\rightarrow u} 
	&= \frac{\hbar}{2} (J_{+\gamma}^{l\rightarrow u} - J_{-\gamma}^{l\rightarrow u}) .
\end{split}\end{alignat}
%-------------------------------------------------------------------------------------------------------------------------

As described in our previous paper\cite{Kitagawa2023},  
we calculate 
the flow $J_{\pm\gamma}^{l\rightarrow u}$ of electrons with spin in the $\pm\gamma$ direction 
by retaining terms up to the second order of $\HT$. 
We assume that the temperature is low enough that 
only a pair of spin-degenerate energy bands $n=0,1$ 
($\veps_{0\vk}^l=\veps_{1\vk}^l$, $\veps_{0\vk}^u=\veps_{1\vk}^u$) contribute to the interlayer electron flow. 
Then $J_{\pm\gamma}^{l\rightarrow u}$ is expressed by 
%-------------------------------------------------------------------------------------------------------------------------
\begin{alignat}{99}\begin{split}\label{eq_J_layer_layer}
	J_{\pm\gamma}^{l\rightarrow u}
	= & \!-\! \frac{2\pi}{\hbar} \!\! \sum_{nn'n''} \sum_{\vk \in B_l} \sum_{\vk' \in B_u} 
	\!\!\!\! \mel{ln\vk}{\HT}{un'\vk'} \!\! \mel{un'\vk'}{\HT}{ln''\vk} \\
		& \times \delta(\veps_{n'\vk'}^u - \veps_{n\vk}^l) \qty(f_{n'\vk'}^u - f_{n\vk}^l) \mel{ln''\vk}{P_{\pm\gamma}}{ln\vk} ,
\end{split}\end{alignat}
%-------------------------------------------------------------------------------------------------------------------------
where $f_{n\vk}^l$ ($f_{n\vk}^u$) is the occupation probability of the lower (upper) sheet.
Owing to the generalized momentum conservation \cite{bistritzer_transport_2010, bistritzer_moire_2011, koshino_interlayer_2015}, 
matrix elements $\mel{ln\vk}{\HT}{un'\vk'}$ and $\mel{un'\vk'}{\HT}{ln''\vk}$ are nonzero 
only when $\vk$ and $\vk'$ sartisfy
%-------------------------------------------------------------------------------------------------------------------------
\begin{alignat}{99}\begin{split}\label{eq_momentum_conservation_layer_layer}
	\vk + \vG_l = \vk' + \vG_u ,
\end{split}\end{alignat}
%-------------------------------------------------------------------------------------------------------------------------
where $\vG_l$ and $\vG_u$ are reciprocal lattice vectors in the lower and upper monolayers, respectively. 
We can limit the sum with respect to $\vG_l$ and $\vG_u$ to those with small absolute values 
since the hopping strength rapidly decays
with increasing $|\vk + \vG_l| \ (=|\vk' + \vG_u|)$ \cite{bistritzer_transport_2010, bistritzer_moire_2011, koshino_interlayer_2015}. 
The momentum conservation Eq.\,\eqref{eq_momentum_conservation_layer_layer} 
reduces  
the expression for $J_{\pm\gamma}^{l\rightarrow u}$ in Eq.\,\eqref{eq_J_layer_layer} 
to the integral with respect to $\vk$, which can be analytically evaluated 
in the case where $f_{n'\vk'}^u - f_{n\vk}^l$ is proportional to $\delta(\veps_{n\vk}^l - \veps_\rF^l)$ 
with $\veps_\rF^l$ the Fermi energy of the lower sheet.  
Such a case will be shown in the subsequent section.   

%==============================================================================================
%\iffigure
\begin{figure}[ht]\centering
	\includegraphics[width=90mm]{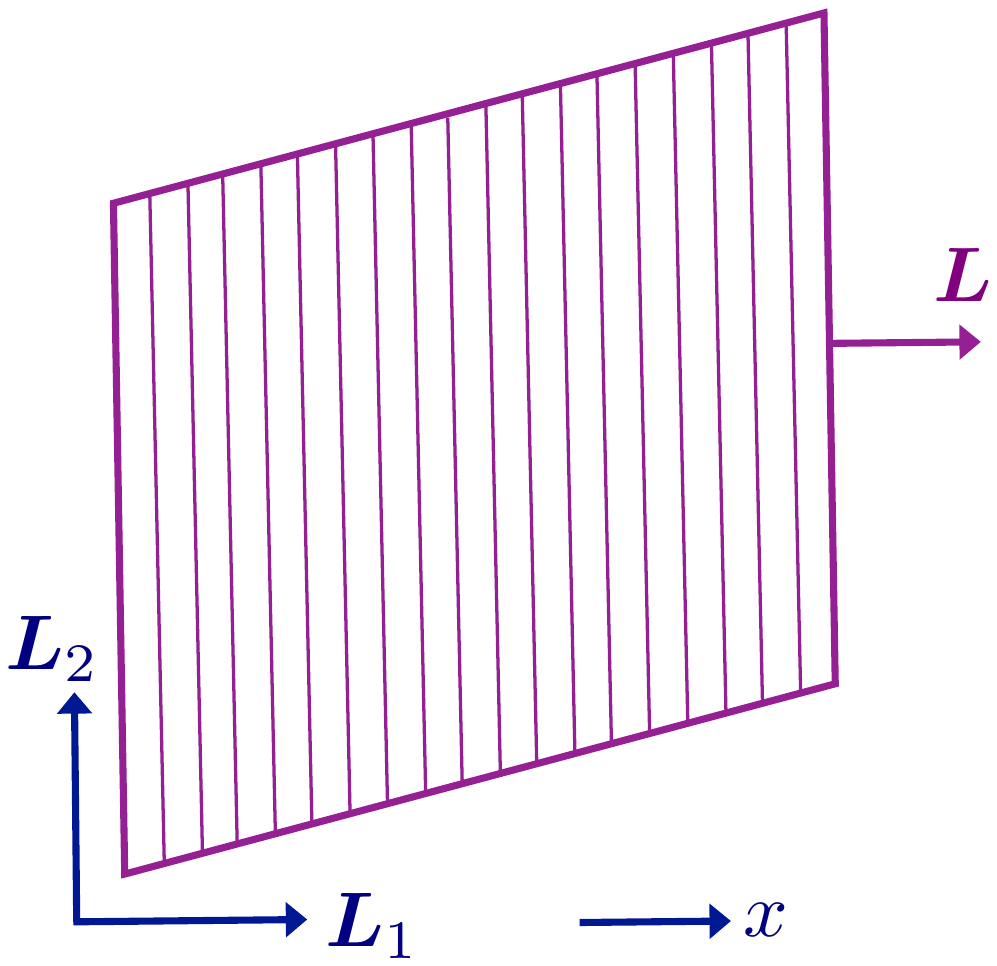}
	\vskip -1.5cm
	\caption{
	The first Brillouin zone of the upper sheet (tube). 
	We choose the parallelogram Brillouin zone with two sides perpendicular to $\vL$.  
	Thin lines perpendicular to $\vL$  
	indicate $\vkappa$ satisfying the periodic boundary condition of the tube, $\vkappa \cdot \vL/(2\pi)=\ $integers.
%	 By choosing the parallelogram Brillouin zone with two sides perpendicular to $\vL$, 
%we can easily count the number of $\vkappa$ points satisfying the periodic boundary conditions of the tube as follows.
%Lines satisying $\vkappa \cdot \vL/(2\pi)=\ $integers are parallel to the side and spaced in the interval of $2\pi/L$ with $L=|\vL|$.
%Since $\vL = m_1 \va_1^u + m_2 \va_2^u$ with $m_1$ and $m_2$ integers, we have $\vb_1^u \cdot \vL = 2\pi m_1$ with $\vb_1^u$ and $\vb_2^u$ the primitive translation vectors in the reciprocal lattice. 
%This means that the distance between two zone boundaries perpendicular to $\vL$ is $m_1$ intervals. 
%Then the number of $\vkappa$ points satisfying the periodic boundary conditions, $\vkappa \cdot \vL/(2\pi)=\ $integers and $\vkappa \cdot \vL_2/(2\pi)=\ $integers, is $(\vb_1^u \cdot \vL) (\vb_2^u \cdot \vL_2)/(2\pi)^2= (L/L_s) N$ with $N = (\vb_1^u \cdot \vL_1) (\vb_2^u \cdot \vL_2)/(2\pi)^2$.  
%Since $N$ is the number of $\vk$ points in the Brillouin zone in the sheet which is equal to the number of unit cells in the sheet, the number of $\vkappa$ points in the tube is confirmed to be the number of unit cells in the tube.  
}
	\label{fig_Brillouin_zone}
\end{figure}
%\fi
%==============================================================================================

%%%%%%%%%%%%%%%%%%%%%%%%%%%%%%%%%%%%%%%%%%%%%%%%%%%%%%%%%%%%%
\subsubsection{Electron flow from an atomic monolayer to a tube}
%%%%%%%%%%%%%%%%%%%%%%%%%%%%%%%%%%%%%%%%%%%%%%%%%%%%%%%%%%%%%%
The electron flow $J_{\pm\gamma}^{l\rightarrow t}$ with spin in the $\pm\gamma$ direction 
from the lower monolayer sheet to the tube is obtained, 
by replacing $\ket{un\vk}$ with $\ket{tn\vkappa}$ in Eq.\,\eqref{eq_J_layer_layer}, to be
%-------------------------------------------------------------------------------------------------------------------------
\begin{alignat}{99}\begin{split}\label{eq_J_tube_layer}
	J_{\pm\gamma}^{l\rightarrow t}
	= & \!-\! \frac{2\pi}{\hbar} \!\!\! \sum_{nn'n''} \sum_{\vk \in B_l} \sum_{\vkappa \in B_u} 
	\!\!\! \mel{ln\vk}{\HT}{tn'\vkappa} \!\! \mel{tn'\vkappa}{\HT}{ln''\vk} \\
		& \times \delta(\veps_{n'\vkappa}^t - \veps_{n\vk}^l) \qty(f_{n'\vkappa}^t - f_{n\vk}^l) \mel{ln''\vk}{P_{\pm\gamma}}{ln\vk} .
\end{split}\end{alignat}
%-------------------------------------------------------------------------------------------------------------------------
We take the momentum summation in the tube as follows.
Since we impose the periodic boundary condition in the direction of $\vL$,  
$\vkappa \cdot \vL$ becomes an integer multiple of $2\pi$ 
and $\vkappa$'s form lines perpendicular to $\vL$ in the two-dimensional momentum space. 
Then it is convenient to take the Brillouin zone of the upper monolayer in the form of parallelogram 
with two sides perpendicular to $\vL$ as shown in Fig.\,\ref{fig_Brillouin_zone}.   
In this Brillouin zone, $\vkappa$ lines are parallel to these sides. 
We take the sum of $\vkappa$ along each of the lines within the Brillouin zone. 

By substituting Eq.\,\eqref{eq_mel_tube_sheet} into tunneling matrix elements in Eq.\,\eqref{eq_J_tube_layer}, 
we have
%-------------------------------------------------------------------------------------------------------------------------
\begin{alignat}{99}\begin{split}
	&\mel{ln\vk}{\HT}{tn'\vkappa}\mel{tn'\vkappa}{\HT}{ln''\vk} = \\
	&\frac{N_s}{N_t} \! \sum_{\vq\vq'} \hat t(\vq) {\hat t}^*\!(\vq') \!
	\mel{ln\vk}{\HT}{un'\vkappa \!+\!\vq} \!\! \mel{un'\vkappa \!+\!\vq'}{\HT}{ln''\vk},
\end{split}\end{alignat}
%-------------------------------------------------------------------------------------------------------------------------
 in which we have the following generalized momentum conservation
%-------------------------------------------------------------------------------------------------------------------------
\begin{alignat}{99}\begin{split}
	\vk + \vG_l = \vkappa + \vq + \vG_u ,\ \ \vk + \vG_l' = \vkappa + \vq' + \vG_u' .
\end{split}\end{alignat}
%-------------------------------------------------------------------------------------------------------------------------
Besides exceptional cases, these equations are satisfied only when 
%-------------------------------------------------------------------------------------------------------------------------
\begin{alignat}{99}\begin{split}\label{eq_momentum_conservation_tube_layer}
	\vk + \vG_l = \vkappa + \vq + \vG_u ,\ \ \vq = \vq',\ \ \vG_l = \vG_l',\ \ \vG_u = \vG_u'. 
\end{split}\end{alignat}
%-------------------------------------------------------------------------------------------------------------------------
Then we obtain 
%-------------------------------------------------------------------------------------------------------------------------
\begin{alignat}{99}\begin{split}\label{eq_matrix_elements_tube_layer}
	&\mel{ln\vk}{\HT}{tn'\vkappa} \mel{tn'\vkappa}{\HT}{ln''\vk} = \\
	& \frac{N_s}{N_t} \sum_{q_x} |\hat t(\vq)|^2  
	\mel{ln\vk}{\HT}{un'\vkappa +\vq} \mel{un'\vkappa +\vq}{\HT}{ln''\vk}.
\end{split}\end{alignat}
%-------------------------------------------------------------------------------------------------------------------------
Here we have taken the $x$ axis in the direction of $\vL$. Then $\vq$ has only the $x$ component 
and $\kappa_x$ becomes an integer multiple of $2\pi/L$ with $L=|\vL|$. 
We can determine $q_x$ and $\kappa_y$ 
in Eq.\,\eqref{eq_J_tube_layer} with Eq.\,\eqref{eq_matrix_elements_tube_layer} by   
the momentum conservation $\vk + \vG_l = \vkappa + \vq + \vG_u$ 
in Eq.\,\eqref{eq_momentum_conservation_tube_layer}. 
Then the expression for $J_{\pm\gamma}^{l\rightarrow t}$ in Eq.\,\eqref{eq_J_tube_layer} 
reduces to the integral with respect to $\vk$, which can be analytically evaluated 
in the case where $f_{n'\vkappa}^t - f_{n\vk}^l$ is proportional to $\delta(\veps_{n\vk}^l - \veps_\rF^l)$   
as shown in the subsequent section 
for the linear-response spin current. 

%==============================================================================================
%\iffigure
\begin{figure}[h]\centering
	\includegraphics[width=95mm]{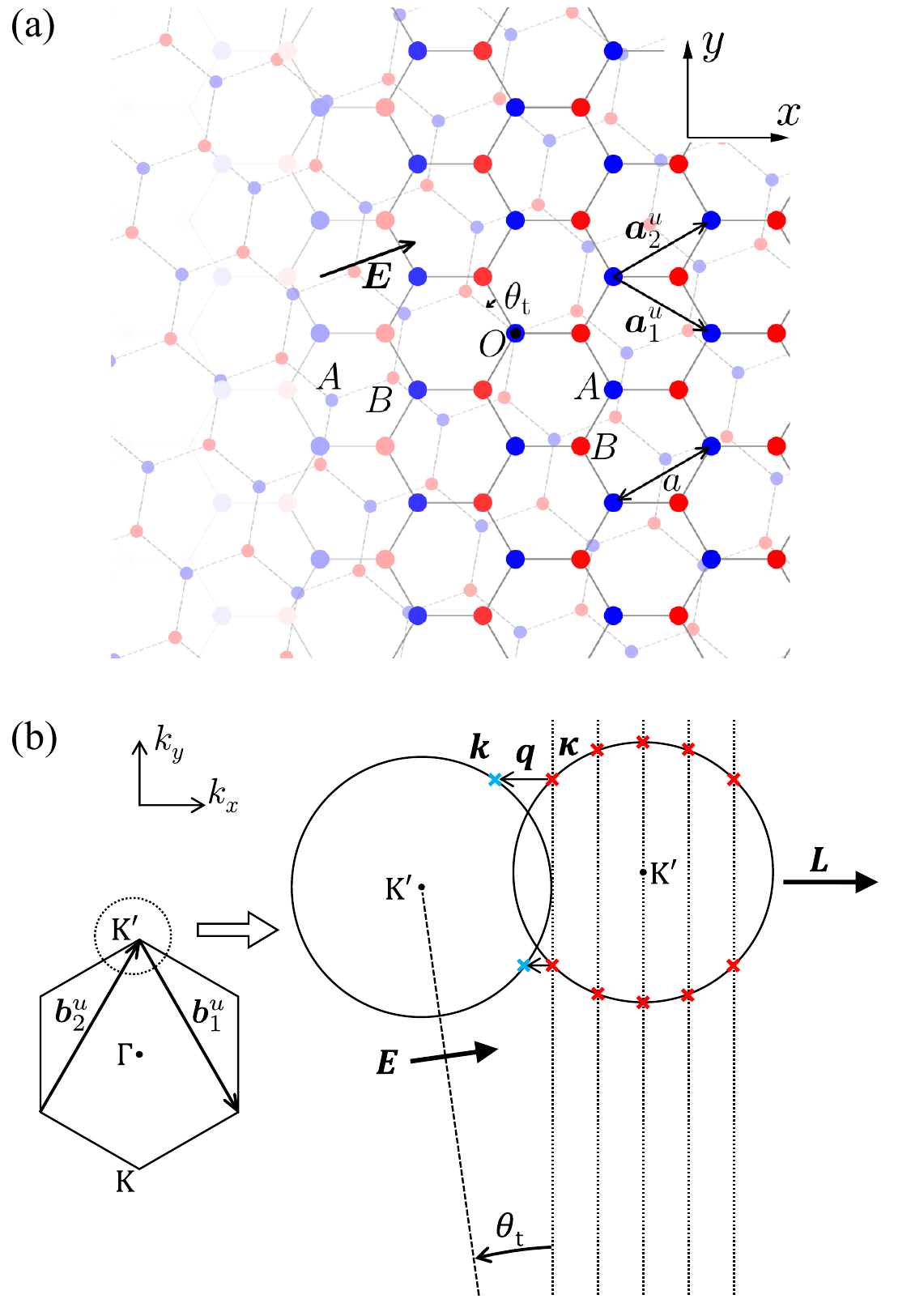}
\vskip -0.8cm
	\caption{
	(a) Top view of the tube-sheet junction of silicene consisting of the armchair tube (solid line shows the junction area of the tube) 
	and the sheet (dashed line). The chiral vector $\vL$ of the tube is in the armchair direction. 
	The sheet is twisted by $\thetat$ with respect to the tube. 
	The electric field $\vE$ to produce the CISP in the sheet is applied in the armchair direction of the sheet. 
	$a$ is the lattice constant.
	(b) The Fermi circle of the twisted sheet and the Fermi points of the tube (red crosses). 
	The blue cross indicates the sheet state at $\vk$ satisfying the momentum conservation 
	for the tunneling to the tube state (red cross) at $\vkappa$
	with the help of the momentum distribution $\vq$ in the tube.}
	\label{fig_silicene_tube_sheet}
\end{figure}
%\fi
%==============================================================================================

%%%%%%%%%%%%%%%%%%%%%%%%%%%%%%%%%%%%%%%%%%%%%%%%%%%%%%%%%%%%%
\section{Spin current from silicene sheet to silicene tube}\label{sec_silicene_junction}
%%%%%%%%%%%%%%%%%%%%%%%%%%%%%%%%%%%%%%%%%%%%%%%%%%%%%%%%%%%%%
As an application of the formula for the sheet-to-tube electron flow Eq.\,\eqref{eq_J_tube_layer}, in this section we calculate 
the spin current from a silicene monolayer sheet to a silicene tube [Fig.\,\ref{fig_silicene_tube_sheet}]. 
Silicene \cite{cahangirov_two_2009, scalise_vibrational_2013, guzman-verri_electronic_2007, vogt_silicene_2012, houssa_silicene_2015, feng_evidence_2012} is one of group-IV atomic layers with the buckled honeycomb structure \cite{balendhran_elemental_2015, molle_buckled_2017}. 
When the current flows in a silicene monolayer, 
staggered CISP is induced in two sublattices $A$ and $B$.  
Owing to the out-of-plane buckling of monolayer silicene, 
the local CISP of sublattice $A$ is extracted more than that of sublattice $B$
by the tube. 
We assume that the tube is in equilibrium with a connected electrode. 

We choose an armchair tube in which $\vL=(L,0)$ is in the armchair direction [Fig.\,\ref{fig_silicene_tube_sheet}(a)]. 
We define the twist angle $\thetat$ by the armchair direction of the lower sheet $\theta_{\ra\rS}$
relative to that of the upper sheet (tube) $\theta_{\ra\rT}$, 
%(in the present armchair tube $\theta_{\ra\rT}=0$)
that is $\thetat=\theta_{\ra\rS} - \theta_{\ra\rT}$. 
We apply the electric field $\vE$ in the armchair direction of the lower sheet.  
As the spin direction $\gamma$ we take 
directions of $\vE$ ($\gamma=\parallel$), 
$\ve_z \times \vE$ ($\gamma=\perp$), 
and $+z$ ($\gamma=z$), 
where $\ve_z$ is the unit vector in the $+z$ direction. 

In the unperturbed Hamiltonian $H_0$ we consider the nearest-neighbor hopping expressed by the Slater-Koster parameter \cite{Slater-Koster1954} 
and take into account the spin-orbit interaction by the LS coupling in each atom. 
We use values of the Slater-Koster parameter and the spin-orbit coupling strength of silicene given in Ref.\,\cite{liu_low-energy_2011}.
Figure\,\ref{fig_silicene_tube_sheet}(b) schematically presents 
the Fermi circle of the sheet and the Fermi points of the tube.  

%==============================================================================================
%\iffigure
\begin{figure}[h]\centering
	\begin{overpic}[width=90mm]{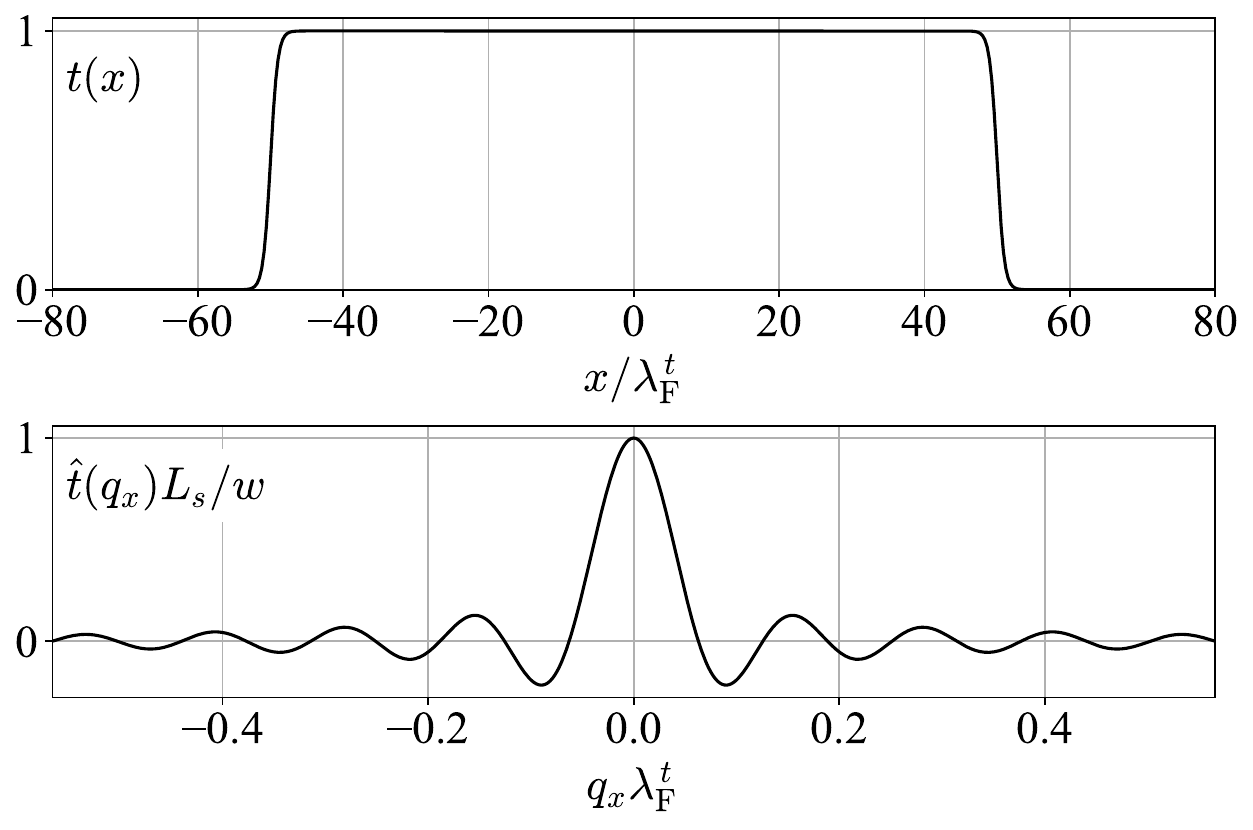}
		\put(-10, 62){(a)}
		\put(-10, 30){(b)}
	\end{overpic} 
	\vskip -0.5cm
	\caption{
	Plots of (a) $t(x)$ [Eq.\,\eqref{eq_t(x)}]  
	(b) $\hat t(q_x)$ [Eq.\,\eqref{eq_t(q_x)}].  
	Values of $w=100\lambda_\rF^t$ and $\lambda=\lambda_\rF^t$ are used.}
	\label{fig_tx_tqx}
\end{figure}
%\fi
%==============================================================================================

In calculating the sheet-to-tube spin current, 
we use $t(x)$ [Eq.\,\eqref{eq_projection_to_junction}] given by
%-------------------------------------------------------------------------------------------------------------------------
\begin{alignat}{99}\begin{split}\label{eq_t(x)}
	t(x)=\frac{1}{2} \left[ \tanh \left( \frac{x+w/2}{\lambda} \right)- \tanh \left( \frac{x-w/2}{\lambda} \right) \right],
\end{split}\end{alignat}
%-------------------------------------------------------------------------------------------------------------------------
where $\lambda$ represents the length scale of variation between $t(x)=0$ and 1. 
Then $\hat t(q_x)$ [Eq.\,\eqref{eq_g_Fourier}] is given, at $L_s \rightarrow \infty$, by
%-------------------------------------------------------------------------------------------------------------------------
\begin{alignat}{99}\begin{split}\label{eq_t(q_x)}
	\hat t(q_x)=\frac{\pi\lambda}{L_s} \frac{\sin(q_x  w/2)}{\sinh(q_x \pi\lambda/2)}. 
\end{split}\end{alignat}
%-------------------------------------------------------------------------------------------------------------------------
Both $t(x)$ and $\hat t(q_x)$ are plotted in Fig.\,\ref{fig_tx_tqx} 
at $w=100\lambda_\rF^t$ and $\lambda=\lambda_\rF^t$. 
These values of $w$ and $\lambda$ are used in the following calculation. 
In Eq.\,\eqref{eq_momentum_conservation_tube_layer}
we take into account three of $\vG_l$ 
which give lower values of $|\vk + \vG_l|$. 
Since the Fermi wavenumber (the radius of the Fermi circle) is much smaller than $|\vG_l|$ and $|\vG_u|$, 
$\vG_u$ satisfying the momentum conservation Eq.\,\eqref{eq_momentum_conservation_tube_layer} 
is only that closest to $\vG_l$. 
We assume that the tube is in equilibrium 
with the temperature $T$ such that $k_\rB T \ll \veps_\rF^t$. 
We obtain the distribution function $f_{n\vk}^l$ in the sheet with the electric field $\vE$ 
by solving the Boltzmann equation in the linear response and in the relaxation-time approximation,  
%-------------------------------------------------------------------------------------------------------------------------
\begin{alignat}{99}\begin{split}\label{eq_boltzmann}
	\frac{-e\vE}{\hbar}\cdot \pdv{f_0(\veps_{n\vk}^l)}{\vk} = -\frac{f_{n\vk}^l - f_0(\veps_{n\vk}^l)}{\tau},
\end{split}\end{alignat}
%-------------------------------------------------------------------------------------------------------------------------
where 
$e\,(>0)$ is the absolute value of the electron charge, 
$f_0(\veps)$ is the Fermi distribution function, and $\tau$ is the constant momentum relaxation time. 
Because deviations of the Fermi surface from a circle are small \cite{Kitagawa2023},
we assume the circular Fermi surface 
and use the linear-in-$k$ dependence of the energy 
in evaluating $\delta(\veps_{n'\vkappa}^t - \veps_{n\vk}^l)$. 
In calculating matrix elements of $\HT$ we use 
the interlayer distance 3.19 \AA\, of bilayer silicene \cite{liu_d_2013} 
and the decay length of the interlayer hopping amplitude 0.184$a$ 
used in the calculation of bilayer graphene \cite{trambly_de_laissardiere_localization_2010, koshino_interlayer_2015}.

%==============================================================================================
%\iffigure
\begin{figure}[ht]
	\includegraphics[width=105mm]{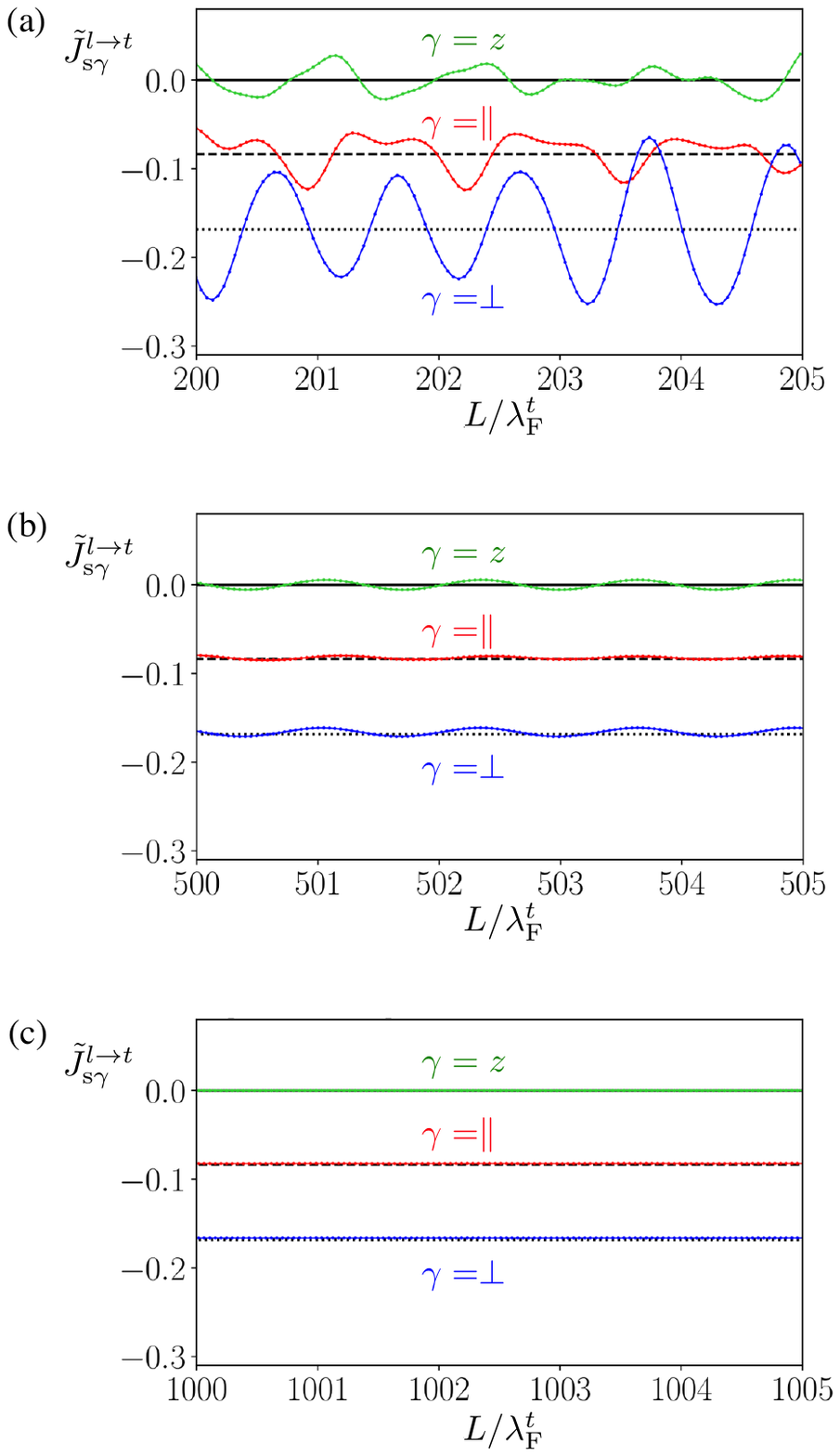}
	\vskip -1.cm
	\caption{	Spin current from silicene sheet to silicene tube 
	$\tilde{J}_{\rs \gamma}^{l\rightarrow t}=J_{\rs \gamma}^{l\rightarrow t}/J_0$ as a function of the tube circumferential length $L$. 
	$J_0= L_s w \tau eE k_{0} V_{pp\pi}^0 / (4\pi\hbar)$ where  
	$V_{pp\pi}^0$ is $|V_{pp\pi}|$ between the nearest neighbor atoms in monolayer silicene 
	and $k_{0} = 0.02 K$ with $K$ the distance between the K and $\Gamma$ points.
	The value of the Fermi wavenumber is $k_{\rF}^l = 0.02K$ in the sheet and $k_{\rF}^t = 0.04K$ in the tube. 
	The twist angle defined in Fig.\,\ref{fig_silicene_tube_sheet}(a) is chosen to be $\thetat=3^{\circ}$. 
	The tunneling-intensity distribution is presented in Fig.\,\ref{fig_tx_tqx}.
	Solid, dashed, and dotted black lines indicate values of the sheet-to-sheet spin current per unit area.
	}
	\label{fig_spin_current_tube_sheet}
\end{figure}
%\fi
%==============================================================================================
%==============================================================================================
%\iffigure
%\begin{figure}[ht]\centering
%	\includegraphics[width=110mm]{fig_spin_current_tube_sheet_components.pdf}
%	\vskip -0.8cm
%	\caption{	Decomposition of the spin current into contributions from three $\vG_l$. $\xi=-1$ in K valley and $\xi=1$ in K$'$ valley.	}
%	\label{fig_spin_current_tube_sheet_components}
%\end{figure}
%\fi
%==============================================================================================

Figure\,\ref{fig_spin_current_tube_sheet} presents 
the spin current from silicene sheet to silicene tube, 
$J_{\rs \gamma}^{l\rightarrow t} 
= (\hbar/2) (J_{+\gamma}^{l\rightarrow t} - J_{-\gamma}^{l\rightarrow t})$ 
($\gamma=\parallel,\perp,z$),
calculated using Eq.\,\eqref{eq_J_tube_layer}. 
Here we place the Fermi level in the conduction band in both the sheet and the tube. 
The Fermi wavenumber is chosen to be 
$k_{\rF}^l = 0.02K$ in the sheet and $k_{\rF}^t = 0.04K$ in the tube, 
where $K$ is the distance between the K and $\Gamma$ points. 
The electron density at $\kF = 0.04 K$ is $6\times 10^{12}$ cm$^{-2}$,
which can be reached in a typical graphene experiment \cite{das_sarma_electronic_2011}. 
We calculate the spin current with increasing the tube circumferential length $L$ 
at a fixed value of the junction width $w=100\lambda_\rF^t$.

We find in plots of $J_{\rs \gamma}^{l\rightarrow t}$ 
for $200<L/\lambda_\rF^t<205$ [Fig.\,\ref{fig_spin_current_tube_sheet}(a)]
that the spin current of each spin direction exhibits an oscillation as a function of $L$. 
The oscillation is quasi-periodic because tunneling processes, which occur in different locations of the momentum space, 
produce oscillations with different periods. 
Each momentum-space location is 
the vicinity of one of the crossing points between upper- and lower-layer Fermi circles.  
Each oscillation period is given by $2\pi/k_\textrm{cross}$ 
where $k_\textrm{cross}$ is the momentum-space distance of the Fermi-circle crossing point 
to the line which is perpendicular to $\vL$ and passes through the K$'$ point 
[a dotted line in Fig.\,\ref{fig_silicene_tube_sheet}(b)]. 
The K valley gives the same contribution to the spin current as that from the K$'$ valley because of the time-reversal symmetry. 
The contribution from each crossing-point vicinity to the spin current
oscillates with $L$ because 
the Fermi points of the tube move along the circle with increasing $L$ 
and cross the Fermi circle of the sheet [Fig.\,\ref{fig_silicene_tube_sheet}(b)].  
In Fig.\,\ref{fig_spin_current_tube_sheet}(a) we also notice that 
the $z$ component in the tube-sheet junction $J_{\rs z}^{l\rightarrow t}$ is nonzero 
in contrast to the sheet-sheet junction 
in which the $C_3$ symmetry leads to $J_{\rs z}^{l\rightarrow u}=0$.
This component, which is allowed to appear when the $C_3$ symmetry is broken,  
inevitably appears because the sum of contributions oscillating with different periods cannot be zero.

Plots of $J_{\rs \gamma}^{l\rightarrow t}$ 
for larger $L$ [Fig.\,\ref{fig_spin_current_tube_sheet} (b) and (c)]
show a decay of the oscillation with increasing $L$.  
This is because the separation between the quantized momenta, $2\pi/L$, 
becomes smaller than 
the momentum uncertainty, $2\pi/w$, given by the width of $\hat t(q_x)$ [Fig.\,\ref{fig_tx_tqx}(b)].
We confirm that the value of the spin current in the tube-sheet junction for each spin direction 
approaches that in the sheet-sheet junction 
as the oscillation decays.

%%%%%%%%%%%%%%%%%%%%%%%%%%%%%%%%%%%%%%%%%%%%%%%%%%%%%%%%%%%%%
\section{Conclusions}\label{sec_conclusion}
%%%%%%%%%%%%%%%%%%%%%%%%%%%%%%%%%%%%%%%%%%%%%%%%%%%%%%%%%%%%%
We have derived an approximate formula for 
tunneling matrix elements of a tube-sheet junction of atomic monolayer, 
expressed with those of the corresponding sheet-sheet junction which have been expressed 
in previous theories \cite{bistritzer_transport_2010, bistritzer_moire_2011, koshino_interlayer_2015} 
with interlayer hopping integrals between atoms. 
The present approximation is applicable to the cases where the width of the tube-sheet junction is much larger than 
the tube Fermi wavelength. 
With use of this formula, we have derived the formula for the electron flow through the junction. 
By applying the derived formula, 
we have calculated the spin current from a silicene sheet with the sublattice-staggered CISP to a silicene tube.  
We have found that the spin current exhibits a quasi-periodic oscillation 
with increasing the tube circumferential length 
due to different-period oscillations in tunneling processes, 
which occur in different locations of the momentum space.  
The contribution from each tunneling process to the spin current oscillates with a constant period 
as the tube Fermi points cross the sheet Fermi circle. 
We have also found that 
the spin current with out-of-plane spin direction, 
which is allowed to appear due to the broken $C_3$ symmetry in the tube-sheet junction, 
appears in the form of oscillation. 
This appearance is inevitable because 
the sum of oscillations with different periods cannot be zero.

% Specify following sections are appendices. Use \appendix* if there only one appendix.
%\appendix
%\section{}

\begin{acknowledgments}
This work was partly supported by Grant-in-Aid for Scientific Research (C) Grant No.
JP21K03413 from the Japan Society for the Promotion of Science (JSPS).
\end{acknowledgments}

% Create the reference section using BibTeX:
%\bibliography{Spintronics_v2_kitagawa}

%%%%% bibliography is included in the main tex file.

\end{document}